\documentclass[aps,twocolumn,nofootinbib,
preprintnumbers]{revtex4-1}
\usepackage{graphicx}
\usepackage{bm}
\usepackage{hyperref}
\usepackage{amsmath,amssymb,amsfonts,graphicx,float}
\renewcommand{\vec}[1]{{\mathbf{#1}}}
\newcommand{\beq}{\begin{eqnarray}}
\newcommand{\eeq}{\end{eqnarray}}
\usepackage{amsmath}
\usepackage{amssymb}
\usepackage[paper=letterpaper,margin=1in]{geometry}
\usepackage{braket}
\usepackage{upgreek}
\newcommand{\abs}[1]{\ensuremath\left\lvert #1 \right\rvert}
\newcommand{\HH}{\ensuremath\mathcal{H}}
\newcommand{\EE}{\ensuremath\mathcal{E}}
\renewcommand{\vec}[1]{\boldsymbol{#1}}
\newcommand{\p}{\ensuremath\uppi}
\newcommand{\ee}{\ensuremath\mathrm{e}}
\newcommand{\dd}{\ensuremath\mathrm{d}}
\newcommand{\ii}{\ensuremath\mathrm{i}}
\DeclareMathOperator{\sgn}{sgn}
\newcommand{\comb}[2]{\begin{pmatrix}#1 \\ #2\end{pmatrix}}
\usepackage{tikz}

\begin{document}

\title{Absence of Luttinger's Theorem due to Zeros in the
  Single-Particle Green Function}

\author{Kiaran B. Dave and Philip W. Phillips}

\affiliation{Department of Physics,
University of Illinois
1110 W. Green Street, Urbana, IL 61801, U.S.A.}
\author{Charles L. Kane}

\affiliation{Department of Physics,
University of Pennsylvania and Astronomy,
Philadelphia, PA 19104, U. S. A. }

\date{\today}

\begin{abstract}
We show exactly with an $SU(N)$ interacting model that even if the ambiguity
associated with the placement of the chemical potential, $\mu$, for a $T=0$
gapped system is removed by using the unique value
$\mu(T\rightarrow 0)$,
Luttinger's sum rule is violated even if the ground-state degeneracy
is lifted by an infinitesimal hopping.  The failure stems from the non-existence of the Luttinger-Ward functional for a system in which the self-energy diverges.  Since it is 
the existence of the Luttinger-Ward functional that is the basis for
Luttinger's theorem which relates the charge density to sign changes
of the single-particle Green function, no such theorem exists.
Experimental data on the cuprates  are presented which show a
systematic deviation from the Luttinger count, implying a breakdown of
the electron quasiparticle picture in strongly correlated electron matter.
\end{abstract}

\maketitle

While the charge density remains fixed under renormalization from high
(UV) to low (IR) energy, precisely what is carrying the charge can change drastically.   For example, in QCD,
free quarks at UV scales form bound
states in the IR.  The key signature of bound quark states is that the
pole in the propagator is converted to a zero\cite{thooft}, implying the fields in the UV-complete theory no longer propagate at low
energy, hence, a breakdown of the elemental particle picture. 
The
conversion of poles to zeros of the single-particle Green function
also obtains in
superconductivity.  In both QCD and superconductivity, the new strongly coupled
ground state in the IR lacks adiabatic
continuity with the UV-state: free quarks (QCD) or free electrons (superconductivity).  The
question of how to compute the number of low-energy charged particle
states is then problematic because what was a particle (pole) at
high energy is no longer so at low energy.   It is not surprising then that the
only
`theorem', due to Luttinger\cite{luttinger}, on the particle density in a fermionic
system, makes no distinction between zeros and poles.  The precise
mathematical statement of Luttinger's theorem for 
spin-$\tfrac{1}{2}$ fermions,
\beq\label{lutt}
n=2\sum_{\vec k}\Theta({\rm Re} G(\vec k,\omega=0)),
\eeq
sums all momenta, $\vec k$, where the Heaviside step function,
$\Theta({\rm Re} G(\vec k,\omega=0)$ is non-zero, with $G(\vec k,\omega)$ the
 time-ordered single-particle Green function.  The right-hand side of
 Eq. (\ref{lutt}) requires ${\rm Re} G(\vec k,\omega=0)=0$ or 
 ${\rm Re} G(\vec k,\omega=0)>0$. The latter obtains either from a pole or a
 zero of the single-particle Green function.  Hence,
 as far as the mathematics is
 concerned, poles and  zeros of the Green function
enter the particle density on the same footing.  We show that
any statement of this kind in which zeros and poles are treated on
equal footing is in error, hence the title of this paper. 
 
While poles of the single-particle Green function represent
quasiparticles, zeros\cite{zeros1} are quite different as they
indicate the presence of a gap\footnote{The actual zero
  condition, ${\rm Det}[G]=0$, ensures that zeros are absent
if adiabaticity with a band insulator is present, e. g. mean-field ordered
  states, (see Appendix).}.  Equivalently, the self energy
 diverges, thereby representing a breakdown of perturbation theory.  As a result, purported non-perturbative proofs of Luttinger's theorem which assume gapless phases at the outset\cite{oshikawa,praz}  bear no relevance to the validity of Eq. (\ref{lutt}).  Rather, such proofs are relevant only to the physical assertion that the volume of the Fermi surface is independent of the interactions---on some level a tautology, since all renormalizations from short-ranged repulsive interactions\cite{polchinski,benefatto,shenkar} are towards the Fermi surface.   
 
Naively, for  gapped incompressible phases at $T=0$, the chemical
potential can be placed anywhere in the gap with impunity.  However,
as first pointed out by Rosch\cite{rosch} for a Mott insulator, the placement of the chemical potential will change the energy at
which $G$ vanishes and hence affect the particle density.  Nonetheless,  
Farid\cite{farid} has argued that the problem arising from this degree
of freedom is entirely spurious because the chemical potential even at
$T=0$ is unique, namely the limiting value of $\mu$ as $T\rightarrow
0$. For the case of the atomic limit of the $\mathrm{SU}(2)$ Hubbard
model, this limiting procedure places the chemical potential equally
far from both bands, the particle-hole symmetric point, and
Eq. (\ref{lutt}) reduces exactly to $n=2\Theta(0)=1$, a result which
holds beyond the atomic limit\cite{stanescu}.  Farid's claim is
interesting then because it would seem to establish a rigorous
relationship between a quantity which has no obvious physical import
and a conserved one, the particle density. In fact, Rosch\cite{rosch,roschresponse}
has shown that Farid's argument, at least perturbatively in the
hopping, is false for a Mott insulator.  What would be advantageous is
a proof which does not rely on perturbation theory and a general
demonstration of precisely where Luttinger's proof goes awry.

We show here exactly using an $SU(N)$ generalization of the atomic
limit of the Hubbard model in which $N$ flavors of ``iso-spin'' and $n$
fermions reside on each site that Farid's\cite{farid}
placement of the chemical potential does not salvage
Eq. (\ref{lutt}). The key result is quite simple. For this model, Eq. (\ref{lutt}) reduces to
\beq\label{luttb}
n=N\Theta(2n-N)
\eeq
which is clearly not equal to the particle density unless $n\in
\{0,N/2,N\}$.   Namely, any partially filled band with $N$ odd leads
to a violation of Eq. (\ref{lutt}). That
Eq. (\ref{luttb}) actually reduces to the correct result for the $\mathrm{SU}(2)$ case is entirely an accident because the $\Theta$ function only takes on values 
of $0$, $1/2$, or $1$.  The crux of the problem is that the
Luttinger-Ward (LW) functional strictly does not exist when zeros of
the Green function are present.  Since
Eq. (\ref{lutt}) relies explicitly on the construction of the LW functional and it does not exist for zeros 
of the Green function, Luttinger's theorem (Eq. (\ref{lutt})) does not exist.
\begin{figure}[h!]
\begin{center}
\includegraphics[width=2.0in]{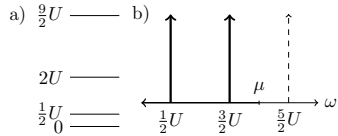}
\caption{  Schematic of the a) energy levels and b) spectral function for the Hamiltonian $H=\tfrac{U}{2}(\hat{n})^2$.  For $n=2$,  $\epsilon_+=H(n+1)-H(n)=5U/2$, $\epsilon_-=H(n)-H(n-1)=3U/2$, and as a consequence, $\mu=(\epsilon_++\epsilon_-)/2=2U$.}
\label{levels}
\end{center}
\end{figure}

To illustrate the problem zeros pose for Eq. (\ref{lutt}), we consider
for simplicity the $SU(N)$ generalization,
\beq
H=\tfrac{U}{2} (n_1+n_2+\cdots+n_N)^2=\tfrac{U}{2}(\hat{n})^2
\eeq
of the atomic limit of the Hubbard model.  Since our key result hinges
only on the existence of zeros, not on the details of a specific
model, our conclusion is general. 
We have not included the
site index here as it is superfluous to the many-body physics which is
captured entirely by the N-flavors of ``iso-spin'' that live on a
single site.  Fig. (\ref{levels}) illustrates the level
structure for $N=3$.   We define $K=H(\hat{n})-\mu \hat{n}$ and write the Green function,
 \beq
G_{\alpha\beta}(\omega)=\frac{1}{Z}\sum_{ab}\ee^{-\beta K_a}Q^{ab}_{\alpha\beta},
\eeq
using the K\"all\'en--Lehmann representation,
where
\beq\label{kallen}
Q^{ab}_{\alpha\beta}=\frac{\braket{a|c_{\alpha}|b}\braket{b|c^\dagger_{\beta}|a}}{\omega-K_b+K_a}+ 
\frac{\braket{a|c^\dagger_{\beta}|b}\braket{b|c_{\alpha}|a}}{\omega-K_a+K_b}
\eeq
in which the Green function is a sum of particle addition and removal parts.
Here $Z$ is the partition function and $K_a=\braket{a|K|a}$.  Since $K_a$ is completely determined by the occupancy number 
$n_a=\sum_{\alpha}\braket{a|c^\dagger_{\alpha}c_{\alpha}|a}$ of the $\mathrm{SU}(N)$ orbital, we may write $K_a=K(n)$ for $n_a=n$.  Noting that 
there are $\begin{pmatrix}N\\n\end{pmatrix}$ states of occupancy number $n$ allows us to simplify the Green function to
\beq
G_{\alpha\beta}(\omega)=\sum_{n=0}^Np(n)Q^{n}_{\alpha\beta}
\eeq
where
\beq
Q^n_{\alpha\beta}=\frac{x_{+\alpha\beta}(n)}{\omega-K(n+1)+K(n)}+\frac{x_{-\alpha\beta}(n)}{\omega-K(n)+K(n-1)},\nonumber
\eeq
 \beq
 p(n)&=\frac{1}{Z}
\comb{N}{n}\ee^{-\beta K(n)},\quad Z=\sum \comb{N}{n}\ee^{-\beta K(n)},\nonumber
\eeq
and the spectral weights are
\begin{subequations}
\beq
x_{+\alpha\beta}(n)&=\comb{N}{n}^{-1}\sum_{n_a=n}\braket{a|c_{\alpha}c^\dagger_{\beta}|a} 
\eeq
for particle addition and
\beq
x_{-\alpha\beta}(n)=\comb{N}{n}^{-1}\sum_{n_a=n}\braket{a|c^\dagger_{\beta}c_{\alpha}|a}
\eeq
\end{subequations}
for particle removal.

To simplify the Green function, it suffices to calculate the spectral weights, $x_{\pm\alpha\beta}$.   Note that in order for these matrix elements to be nonvanishing, one must have $\alpha=\beta$. In addition, the state $a$ occurring in the summation expression for $x_{+\alpha\beta}$ (resp. $x_{-\alpha\beta}$) must be empty (full) at isospin $\alpha$. There are $\comb{N-1}{n}$ ($\comb{N-1}{n-1}$) such states, and so the final expressions for $x_{\pm\alpha \beta}$ are
\begin{subequations}
\label{x}
\beq\label{x+}
x_{+\alpha\beta}&=\delta_{\alpha\beta}\comb{N-1}{n}/\comb{N}{n}=\delta_{\alpha\beta}(1-\frac{n}{N})
\eeq
and
\beq\label{x-}
x_{-\alpha\beta}&=\delta_{\alpha\beta}\comb{N-1}{n-1}/\comb{N}{n}=\delta_{\alpha\beta}\frac{n}{N}.
\eeq
\end{subequations}
At $T=0$, $p(n)=1$ for some fixed $n$ and $p(n)=0$ otherwise which effectively eliminates the sum over $n$ in the Green function resulting
in an expression of the form, $G_{\alpha\beta}(\omega)=\delta_{\alpha\beta}Q^n$, where $Q^n(\omega)$ is $Q^n_{\alpha\beta}(\omega)$ evaluated with
Eq. (\ref{x}).  We now come to the issue of the chemical potential. According to Farid\cite{farid}, the leading $\beta\rightarrow \infty$ limit of the Green function is given by evaluating
\beq
&Q&^n(\omega)+\ee^{-\beta(K(n+1)-K(n))}(Q^{n+1}(\omega)-Q^n(\omega))\nonumber\\
&+&\ee^{\beta(K(n)-K(n-1))}(Q^{n-1}(\omega)-Q^n(\omega))\nonumber\\
&+&O(\ee^{-\beta(H(n+1)-2H(n)+H(n-1))}).
\eeq
The chemical potential is fixed by the relationship
\beq
\frac{n}{N}=\lim_{\beta\rightarrow \infty}\int \frac{\dd \omega}{\ee^{\beta\omega}+1}\left(-\frac{1}{\p}\Im G(\omega+\ii 0)\right).
\eeq
Combining the previous two expressions yields
\beq
\frac{n}{N}&=&\left(\tfrac{n}{N}\right)(1-\ee^{-\beta(K(n+1)-K(n)})\nonumber\\
&+&\left(1-\tfrac{n}{N}\right)e^{\beta(K(n)-K(n-1))}\nonumber\\
&+&O(\ee^{-\beta(H(n+1)-2H(n)+H(n-1))}),
\eeq
which can be solved immediately to yield
\beq
\mu=\frac{\epsilon_++\epsilon_-}{2}+\frac{1}{2\beta}\log \frac{N-n}{n}+o(\beta^{-1})
\eeq
where $\epsilon_+(n)=H(n+1)-H(n)$ and $\epsilon_-(n)=H(n)-H(n-1)$. Consequently, the chemical potential is equidistant between the poles of $Q^n(\omega)$.  
Equivalently, this choice for the chemical potential implies that $K(n+1)-K(n)>0$, $K(n)-K(n-1)<0$, and $K(n+1)-K(n)=-(K(n)-K(n-1))$. Hence
\beq
G_{\alpha\beta}(\omega=0)=\frac{\delta_{\alpha\beta}}{K(n+1)-K(n)}\left(\frac{2n-N}{N}\right).
\eeq
Consequently, Luttinger's theorem for this system, if it is valid, is the statement that
\beq\label{thetan}
n=\sum_{\alpha}\Theta(G_{\alpha\alpha}(\omega=0))=N\Theta(2n-N).
\eeq
This expression clearly fails for any partial filling when $N$ is
odd. Also, by making the shift $\omega\rightarrow\omega+\epsilon(\vec
k))$ in Eq. (\ref{kallen}), thereby lifting the ground-state
degeneracy (see Appendix) cannot change the argument of
the $\Theta$-function in Eq. (\ref{thetan}) and the criticism of Ref. [13] fails.  Consider the atomic limit of the Hubbard model for $N=3$ and two of the iso-spin levels with unit occupancy, that is, $n=2$ (see Fig. 1).  This is the `half-filled' case.   Eq. (\ref{luttb}) clearly fails because $\Theta(x)$ can only take on values $1$, $0$ or $1/2$.  Hence, no expression of the form of Eq. (\ref{luttb}) can ever yield the electron density when $N$ is odd.   At play here is the fact that particle-hole symmetry, which is present for $N$ even,
 is strictly absent for $N$ odd.  

Clearly if Eq. (\ref{luttb}) fails, there must be an additional term that contributes to the density.  The extra term is usually\cite{luttinger}
written as an integral involving the self energy.
As $\beta \rightarrow \infty$, $G(\omega)\equiv G_{\alpha\alpha}(\omega)$ may be rewritten as
\beq
G(\omega)=\frac{1}{\omega+\mu-\bar{\epsilon}-\Sigma(\omega) }
\eeq
where
\beq
\bar{\epsilon}&=\left(1-\frac{n}{N}\right)\epsilon_++\left(\frac{n}{N}\right)\epsilon_-,
\eeq
\beq\label{sigma}
\Sigma(\omega)&=\frac{n(N-n)}{N^2}\frac{(\epsilon_+-\epsilon_-)^2}{\omega+\mu-\epsilon_0}
\eeq
and 
\beq
\epsilon_0&=\left(\frac{n}{N}\right)\epsilon_++\left(1-\frac{n}{N}\right)\epsilon_-.
\eeq
The expression to be evaluated is  
\beq
I_2=-N\lim_{\beta\rightarrow \infty} \frac{1}{\beta}\sum_{\zeta}\left. G(\omega)\partial_{\omega}\Sigma(\omega)\right\rvert_{\omega=\ii\zeta}
\eeq
where the sum is over the fermionic Matsubara frequencies.  The explicit calculation yields\cite{farid}
\beq
I_2&=&\frac{N}{2}\lim_{\beta\rightarrow\infty}\left(\left(\tfrac{n}{N}\right)\tanh\beta(\epsilon_+-\mu)\right.\\
&\quad+&\left.\left(1-\tfrac{n}{N}\right)\tanh\beta(\epsilon_--\mu)-\tanh\beta(\epsilon_0-\mu) \right)\nonumber\\
&=&\frac{1}{2}\left(n \sgn (\epsilon_+-\mu)+(N-n)\sgn(\epsilon_--\mu)\right.\nonumber\\
&\quad-&\left.N\sgn(\epsilon_0-\mu)\right).\nonumber
\eeq
 Because $\mu=\frac{\epsilon_-+\epsilon_+}{2}$ and $\epsilon_-<\epsilon_+$, we know that $\epsilon_-<\mu$, $\epsilon_+>\mu$, and $\sgn(\epsilon_0-\mu)=\sgn(2n-N)$. Thus,
\beq
I_2&=&\frac{1}{2}(2n-N-N\sgn(2n-N))\nonumber\\&=&n-N\Theta(2n-N).
\eeq
Combined with the previous result, $I_1=N\Theta(2n-N)$, we recover the full particle density,
\beq
n=I_1+I_2.
\eeq

The failure of the LW identity, $I_2=0$, rests entirely on the form of the self-energy in this problem,  
Eq. (\ref{sigma}).  We first note that $\Sigma$ diverges at $\omega+\mu=\epsilon_0$.  
Consequently, regardless of what is chosen for $H_0$, which in this case has been set to zero, $\Sigma$ cannot be 
connected to any non-interacting problem as a result of its divergence.  Consider the LW functional, defined by
\begin{subequations}
\begin{align}
\delta I[G]&=\int d\omega \Sigma\delta G\\
I[G=G_0]&=0
\end{align}
\end{subequations}
which was used by Luttinger\cite{luttinger} to show that the integrand
of $I_2$ is a total derivative. Because $\Sigma$ diverges for some
$\omega$ when $G$ is the total Green function, it is not possible to
integrate the defining differential expression in the neighborhood of
the true Green function, and therefore the LW functional does not
exist.  Consequently, there is no Luttinger theorem and
Eq. (\ref{lutt}) does not represent the density of a fermionic system
because zeros of the Green function must be strictly excluded, a
model-independent conclusion.  While zeros are a robust mathematical feature of a Green function, they
do not represent a conserved quantity and do not have direct physical meaning.
This can be seen from the fact that the zero line is sensitive to the placement of the
chemical potential, and there is no observable consequence when
the zero crosses $\mu$. 

Even in the case where $I_2 = 0$, and our system is gapped, the
divergence of the self-energy is still present.   Hence,
no LW functional exists in this case as well.
For N even, particle-hole symmetry is operative and it
is this symmetry that results in a vanishing of $I_2$
not any fact pertaining to the LW functional. For N
odd, no such symmetry obtains.  In gapless systems,
Eq. (1) is still not generally valid.   A less general formulation\cite{horava,volovik}
which {\it assumes} the absence of zeros remains valid.  That
assumption, means that the interacting system must be perturbatively (adiabatically)
connected to non-interacting fermions, which immediately rules out the
Mott state which has a divergent self energy.  In fact, the work by Ho\v rava\cite{horava}
provides a promising direction in which the robust features of a Fermi surface
admit a K-theoretic formulation.  

The key implication the inapplicability of Eq. (\ref{lutt}) portends for strongly interacting
electron systems is that
although the degrees of freedom that give rise to zeros undoubtedly
contribute to the current, they have no bearing on the particle
density.  The particle density is determined by coherence ($\Im\Sigma<\epsilon$) while zeros
arise from incoherence ( $\Im\Sigma$ diverges\cite{stanescu} signifying that
there is no particle to contribute to $n$).   As a result, there exist charged degrees of
freedom in strongly correlated electron matter which couple to the
current but nonetheless cannot be given an interpretation in terms of
elemental fields.   Consequently, the particle density will be less
than the total number of degrees of freedom that couple to an external
gauge field as demonstrated recently\cite{hong} when the upper Hubbard
band is integrated out exactly. Note the breakdown of Eq. (\ref{lutt})
has been demonstrated exactly in a model in which spin and charge are
not separated, a purely atomic limit model in which there can be no
difference between spin and charge velocities.  

\begin{figure}[h!]
\begin{center}
\includegraphics[width=3.0in]{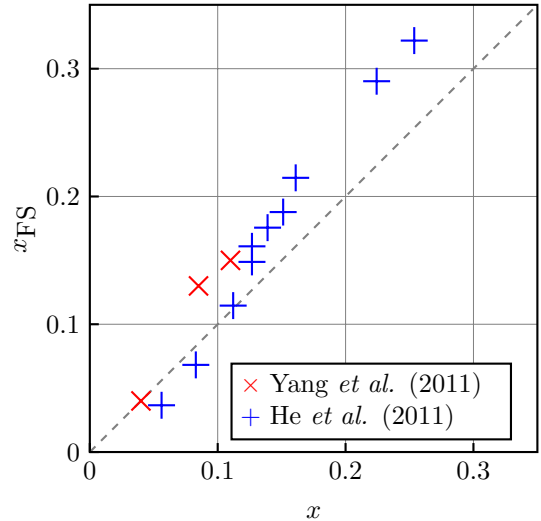}
\caption{Apparent doping $x_{\mbox{FS}}$ inferred from the Fermi surface reconstruction as a function of the nominal doping $x$ in LSCO and Bi-2212. }
\label{fig:ARPESplot}
\end{center}
\end{figure}

Deviations from Eq. (\ref{lutt})  are expected then in experimental
systems which are strongly correlated as evidenced by either a hard gap or a
pseudogap ( density of states vanishing at a single energy).  Shown in
Fig. (\ref{fig:ARPESplot}) is a plot of the area enclosed by the locus of k-points for which there is a maximum
in the spectral function in La$_{2-x}$Sr$_x$CuO$_2$\cite{shen} ($+$
plot symbol) and Bi$_2$Sr$_2$CaCu$_2$O$_{8+\delta}$\cite{kidd}
($\times$ plotting symbol) as a function of the nominal doping level
in the pseudogap regime. Although the maxima in the spectral
function form an arc as there are zeros present on the opposite side,
$x_{\rm FS}$ was extracted by simply closing the arc
according to a recent proposal\cite{yrz} for
Bi$_2$Sr$_2$CaCu$_2$O$_{8+\delta}$\cite{kidd} (Bi-2212) and for
La$_{2-x}$Sr$_x$CuO$_2$\cite{shen} (LSCO) by
determining the large Fermi surface ($1-x$) defined by the $k_F$ measured
directly from the momentum-distribution curves and then
subtracting unity.  Hence, the key
assumption that is being tested here in this definition of $x_{\rm
  FS}$ is that each doped hole corresponds to a single $k-$state.  A typical uncertainty in these experiments
is $\pm 0.02$.  Even when this uncertainty is considered, the
deviation from the dashed line persists indicating that one hole does
not equal one k-state and hence a fundamental breakdown of the elemental
particle picture in the cuprates.  For the Hubbard model this
systematic deviation has been seen previously\cite{mottness} and stems
from the fundamental fact that  since the spectral weight is carried
by two non-rigid bands, removing a single $k$-state is not equivalent
to removing a single electron. Another source of deviation from
Eq. (\ref{lutt}) is the fact that as $x$ nears
optimal doping, the Fermi surface topology must change from
scaling with $x$ to $1-x$.  Hence, there has to be a deviation $x_{\rm
  FS}=x$.   We hope this work serves to motivate a much more
systematic study of deviations from Eq. (\ref{lutt}).

\acknowledgements{ This work was motivated by a vigorous 
  exchange P. W. P. had with Mike Norman over the meaning of zeros as
  part of a joint collaboration funded by the Center for Emergent Superconductivity, a DOE Energy Frontier Research Center, Grant No.~DE-AC0298CH1088. S. Hong, R. Leigh,
  W. Lee, W. Lv, J. Teo and J. Zaanen also provided invaluable questions and comments which shaped this work.  K. Dave and P. Phillips received research financial support from the
NSF-DMR-1104909 and Charles L. Kane is funded by NSF grant
DMR-0906175.}

{\bf Supplemental Material:  Absence of Luttinger's theorem}

\section{Multi-band System}
The derivation of Luttinger's theorem concerns only the
\emph{eigenvalues} of the propagator $G$ on the Hilbert space of
one-particle states\footnote{See, e.g., Dzyaloshinskii (2003).}, not
matrix elements that happen to lie on the diagonal. The irrelevance of
the latter is underlined by the invariance of the entirety of the
proof under the transformation $G\mapsto UGU^{\dagger}$, where $U$ is
an arbitrary unitary transformation on the Hilbert
space. Consequently, methods of projection or integration of degrees
of freedom that do not respect the eigenvalues of the propagator are
incompatible with the familiar statement of the theorem, Eq. (1) in
our manuscript. To illustrate, consider a model of two hybridized bands $a,b$. The Hamiltonian for this system is
\begin{equation}
\begin{split}
H&=\sum_{\vec{k}}\left(\epsilon_{\vec{k}a}a^*_{\vec{k}}a_{\vec{k}}+\epsilon_{\vec{k}b}b^*_{\vec{k}}b_{\vec{k}}
+\Delta_{\vec{k}}a^*_{\vec{k}}b_{\vec{k}}+\Delta^*_{\vec{k}}b^*_{\vec{k}}a_{\vec{k}}\right)\\
&=\sum_{\vec{k}}\psi^{\dagger}_{\vec{k}}\HH_{\vec{k}}\psi_{\vec{k}}
\end{split}
\end{equation}
where
\begin{align}
\psi_{\vec{k}}&=\begin{bmatrix}a_{\vec{k}}\\ b_{\vec{k}}\end{bmatrix}&\HH_{\vec{k}}&=\begin{bmatrix}\epsilon_{\vec{k}a} &\Delta_{\vec{k}}\\ \Delta^*_{\vec{k}} &\epsilon_{\vec{k}b}\end{bmatrix}\mbox{.}
\end{align}
The propagator is then given by the $2\times 2$ matrix
\begin{equation}
\begin{split}
G(\vec{k},\omega)&=\frac{1}{\omega-\HH_{\vec{k}}}\\
&=U_{\vec{k}}\frac{1}{\omega-\EE_{\vec{k}}}U^{\dagger}_{\vec{k}}
\end{split}
\end{equation}
where the diagonalization $\HH_{\vec{k}}=U_{\vec{k}}\EE_{\vec{k}}U^{\dagger}_{\vec{k}}$, $U_{\vec{k}}U^{\dagger}_{\vec{k}}=1$ is given by
\begin{subequations}
\begin{align}
U_{\vec{k}}&=\begin{bmatrix}u_{\vec{k}}&v_{\vec{k}}\\ -v^*_{\vec{k}} & u^*_{\vec{k}}\end{bmatrix}\mbox{,}&\EE_{\vec{k}}&=\begin{bmatrix}\varepsilon_{\vec{k}+}&0\\ 0&\varepsilon_{\vec{k}-}\end{bmatrix}\mbox{,}
\end{align}
\begin{align}
\varepsilon_{\vec{k}\pm}&=&\bar{\epsilon}_{\vec{k}}\pm\sqrt{\epsilon_{\vec{k}}^2+\left\lvert\Delta_{\vec{k}}\right\rvert^2},\quad\bar{\epsilon}_{\vec{k}} =\frac{\epsilon_{\vec{k}a}+\epsilon_{\vec{k}b} }{2}\mbox{,}\nonumber\\\epsilon_{\vec{k}}&=&\frac{\epsilon_{\vec{k}a}-\epsilon_{\vec{k}b}}{2}\mbox{.}
\end{align}
\end{subequations}
Luttinger's theorem holds for this system:
\begin{equation}
n=\sum_{\vec{k}\pm}\Theta(G_{\pm}(\vec{k},\omega=0))= \sum_{\vec{k}\pm}\Theta(-\varepsilon_{\vec{k}\pm})\mbox{.}
\end{equation}
Note, $\rm Det[G]$ does not have any zeros; consequently any
subsequent manipulations which result in zeros must be spurious.

The $aa$ component of $G$ is
\begin{equation}
G_{aa}(\vec{k},\omega)=\frac{\abs{u_{\vec{k}}}^2}{\omega-\varepsilon_{\vec{k}+}}+\frac{\abs{v_{\vec{k}}}^2}{\omega-\varepsilon_{\vec{k}-}}=\frac{1}{\omega-\epsilon_{\vec{k}a}-\frac{\abs{\Delta_k}^2}{\omega-\epsilon_{\vec{k}b}}}
\end{equation}
and therefore the number density $n_a$ of electrons in band $a$ is
\begin{equation}
n_a=\sum_{\vec{k}}\left(\abs{u_{\vec{k}}}^2\Theta(-\varepsilon_{\vec{k}+})+\abs{v_{\vec{k}}^2}\Theta(-\varepsilon_{\vec{k}-})\right)\mbox{.}
\end{equation}
Suppose that the $b_{\vec{k}}$ were integrated out. A naive substitution into the statement of Luttinger's theorem would yield
\begin{equation}
n_{?}=\sum_{\vec{k}}\Theta(G_{aa}(\vec{k},\omega=0))=\sum_{\vec{k}}\Theta(-\varepsilon_{\vec{k+}}\varepsilon_{\vec{k}-}/\epsilon_{\vec{k}b})\mbox{.}
\end{equation}
The quantity $n_{?}$ is equal neither to $n_a$ nor to $n$. This
failure stems from the fact that the Hamiltonian is off-diagonal in
the $ab$ basis. In a general, non-diagonalizing basis, the
propagator's matrix elements will be superpositions of poles and thus
possess zeros, none of which have anything to do with Luttinger's
theorem.

\section{non-Degenerate Ground States}

We allude to in the text that our proof applies even in the presence
of a vanishingly small hopping matrix element, $t=0^+$, which lifts the
degeneracy of the atomic limit.  We now make this explicit by
considering a Green function
\beq
G_{ab}(\omega)={\rm Tr}\left(\left[c_a^\dagger\frac{1}{\omega-H}c_b+c_b\frac{1}{\omega-H}c_a^\dagger\right]\rho(0^+)\right)\nonumber
\eeq
evaluated at $\mu=0$ in the limit $t\rightarrow 0$ and $T=0$, with $\rm Tr$ the trace over
 the Hilbert space and $\rho(0^+)$ the density matrix.  Our use of $\rho(0^+)$ is crucial here because $\rho(t=0^+)$
 describes a pure state whereas for $\rho(t=0)=\sum_u
 P_u|u\rangle\langle u|$ with probabilities satisfying $\sum_u
 P_u=1$ is a mixture of many
 degenerate ground states $|u\rangle$.  Our use of $\rho(0^+)$ allows
 us to do
 perturbation theory in $t$.  For $t\rightarrow 0$, the intermediate
 states have energy $U$ or $0$ and hence we can safely pull $\omega-H$
 outside the trace.  Noting that $\{c_a^\dagger,c_b\}=\delta_{ab}$, we obtain that
\beq\label{defcon5}
G_{ab}(\omega)=\frac{\omega\delta_{ab}-U\rho_{ab}}{\omega(\omega-U)},
\eeq
where we have introduced $\rho_{ab}={\rm Tr}\left(  c_a^\dagger c_b
  \rho(0^+)\right)   =  \langle u_0 |  c_a^\dagger c_b | u_0 \rangle$,
$|u_0\rangle$ the unique ground state.  

Now comes the crucial point. Consider $N=3$. If a magnetic field were to lift the
degeneracy, only a single one of the iso-spin states would be possible
and $\rho_{ab}(0^+)={\rm diag}(1,0,0)\ne \rho_{ab}(t=0)$.  However, turning on a
hopping matrix element places no restriction on the permissible
iso-spin states.  Consider a rotationally-invariant spin singlet state
on a three-site system, the unique ground state.  
We have that $\rho_{ab}(0^+)=1/3
{\rm diag}(1,1,1)=\rho_{ab}(t=0)$. For this unique
ground state, Eq. (\ref{defcon5}) has a zero at $\omega=U/3$ whereas
 $\lim_{T\rightarrow 0}\mu(T)=U/2$.  Consequently, the Luttinger
count fails to reproduce the particle density. Specifically,
Eq. (14) implies that $1=0$!  The key point is that as long as the perturbation
which lifts the degeneracy does not break $SU(N)$ symmetry, then
$\rho_{ab}(t=0^+)=\rho_{ab}(t=0)$ and Eq. (14)
survives which means that any criticism along the lines of
Ref.[13] does not.

\end{document}